\documentclass[english,titlepage]{article}
\usepackage[T1]{fontenc}
\usepackage[latin9]{inputenc}
\usepackage[letterpaper]{geometry}
\usepackage{mathtools}
\usepackage{amsthm}
\usepackage{amsmath}
\usepackage{amssymb}
\usepackage{graphicx}
\usepackage{esint}
\usepackage{epstopdf}
\usepackage{hyperref}
\usepackage{verbatim}

  \usepackage{color}

\makeatletter
\numberwithin{equation}{section}
\numberwithin{figure}{section}

\usepackage{ae,aecompl}

\allowdisplaybreaks
\numberwithin{equation}{section}

\@ifundefined{showcaptionsetup}{}{%
 \PassOptionsToPackage{caption=false}{subfig}}
\usepackage{subfig}
\makeatother

\usepackage{babel}
\begin{document}
\global\long\def\mA{\mathcal{A}}
\global\long\def\mB{\mathcal{B}}
\global\long\def\mC{\mathcal{C}}
\global\long\def\mD{\mathcal{D}}
\global\long\def\mE{\mathcal{E}}
\global\long\def\mF{\mathcal{F}}
\global\long\def\mG{\mathcal{G}}
\global\long\def\mH{\mathcal{H}}
\global\long\def\mI{\mathcal{I}}
\global\long\def\mJ{\mathcal{J}}
\global\long\def\mK{\mathcal{K}}
\global\long\def\mL{\mathcal{L}}
\global\long\def\mM{\mathcal{M}}
\global\long\def\mN{\mathcal{N}}
\global\long\def\mO{\mathcal{O}}
\global\long\def\mP{\mathcal{P}}
\global\long\def\mQ{\mathcal{Q}}
\global\long\def\mR{\mathcal{R}}
\global\long\def\mS{\mathcal{S}}
\global\long\def\mT{\mathcal{T}}
\global\long\def\mU{\mathcal{U}}
\global\long\def\mW{\mathcal{W}}
\global\long\def\mX{\mathcal{X}}
\global\long\def\mY{\mathcal{Y}}
\global\long\def\mZ{\mathcal{Z}}

\global\long\def\sl#1{\ensuremath{\mathrlap{\!\not{\phantom{#1}}}#1}}
\global\long\def\e{\epsilon}
\global\long\def\ra{\rightarrow}
\global\long\def\la{\leftarrow}
\global\long\def\avg#1{\left\langle #1\right\rangle }

\global\long\def\f#1#2{\frac{#1}{#2}}
\global\long\def\del{\partial}
\global\long\def\t{\theta}
\global\long\def\a{\alpha}
\global\long\def\b{\beta}
\global\long\def\g{\gamma}
\global\long\def\G{\Gamma}
\global\long\def\s{\sigma}
\global\long\def\r{\rho}
\global\long\def\d{\delta}
\global\long\def\Tr{\text{Tr}}
\global\long\def\tr{\text{tr}}
\global\long\def\ket#1{\left\langle #1\right|}
\global\long\def\bra#1{\left|#1\right\rangle }
\global\long\def\N{\mathbb{N}}
\global\long\def\Z{\mathbb{Z}}
\global\long\def\R{\mathbb{R}}
\global\long\def\C{\mathbb{C}}
\global\long\def\p{\varphi}
\global\long\def\msbar{\overline{\text{MS}}}
\global\long\def\T{\text{T}}
\global\long\def\w{\omega}
\global\long\def\D{\Delta}
\global\long\def\O{\Omega}
\global\long\def\S{\Sigma}

\vspace*{\fill}
\begin{center}
\Large \bf{Traversable Wormholes via a Double Trace Deformation}\\
\vspace{.5 in}
\large\bf{ Ping Gao$^1$, Daniel Louis Jafferis$^1$, Aron C. Wall$^2$}\\
\vspace{12 pt}
\large $^1$\it{Center for the Fundamental Laws of Nature, Harvard University, Cambridge, MA, USA}\\
\large $^2$\it{School of Natural Sciences, Institute for Advanced Study,
 Princeton, NJ, USA}\\
\vspace{2 in}
\large\bf Abstract \\
\end{center}

After turning on an interaction that couples the two boundaries of an eternal BTZ black hole, we find a quantum matter stress tensor with negative average null energy, whose gravitational backreaction renders the Einstein-Rosen bridge traversable.  Such a traversable wormhole has an interesting interpretation in the context of ER=EPR, which we suggest might be related to quantum teleportation.  However, it cannot be used to violate causality.  We also discuss the implications for the energy and holographic entropy in the dual CFT description.
\vspace*{\fill}

\newpage{}

\tableofcontents{}

\section{Introduction}

Traversable wormholes have long been a source of fascination as a
method of long distance transportation \cite{Morris1988a}. However,
such configurations require matter that violates the null energy condition,
which is believed to apply in physically reasonable classical theories.
In quantum field theory, the null energy condition is false, but in
many situations there are other no-go theorems that rule out traversable wormholes.

In this work we find that adding certain interactions that couple
the two boundaries of eternal AdS-Schwarzschild results in a quantum
matter stress tensor with negative average null energy, rendering
the wormhole traversable after gravitational backreaction.  The coupling we turn on has the effect of modifying the boundary conditions of a scalar field in the bulk, which changes the metric at 1-loop order.

Violation of the averaged null energy condition (ANEC) is a prerequisite
for all traversable wormholes \cite{Morris1988,Visser1996,Visser2003,Hochberg1998}.
It states that there must be infinite null geodesics  passing through the wormhole, with tangent vector $k^{\mu}$ and affine parameter $\lambda$,
along
which
\begin{equation}
\int^{+\infty}_{-\infty} T_{\mu\nu}k^{\mu}k^{\nu}d\lambda<0.
\end{equation}
The physical picture is that by Raychaudhuri's equation for null geodesic
congruence, light rays will defocus only when ANEC is violated. In
that case, the light rays that focus in one end
of the wormhole can defocus when going out the other end.

There are reasonable arguments that the ANEC is always obeyed along infinite achronal geodesics \cite{Graham:2007va, Kontou:2012ve, Kontou:2015yha, Kelly:2014mra, Wall:2009wi}.\footnote{A set of points is achronal if no two of the points can be connected by a timelike curve; otherwise it is chronal.}  This is sufficient to rule out traversable wormholes joining two otherwise disconnected regions of spacetime \cite{Graham:2007va}.  Furthermore, the generalized second law (GSL) of causal horizons also rules out traversable wormholes connecting two disconnected (asymptotically flat or AdS) regions, due to the fact that the future horizon of a lightray crossing through the wormhole has divergent area at very early times, which contradicts the increase of generalized entropy along the future horizon \cite{Wall2013}.

For small semiclassical perturbations to a stationary causal horizon, both the GSL and the ANEC follow from lightfront quantization methods that are valid for free or superrenormalizable field theories \cite{Wall2012}.  (There is also evidence that these results extend to more general field theories \cite{Faulkner:2016mzt,Hofman:2008ar,Hofman:2016awc,Koeller:2015qmn,Bunting:2015sfa}).

In our configuration, signals from early times on the horizon can intersect it again at late times, by passing through the directly coupled boundaries.  The causal structure of the manifold is modified as a result, changing the commutation relations along null geodesics through the wormhole and making them no longer achronal.  For the same reason, a causal horizon extending through the wormhole intersects itself, removing the piece with divergent area.  Hence the above impossibility results do not apply.  The negative energy matter in our configuration is similar to the Casimir effect, since the interaction between the boundaries implies that the radial direction is effectively a compact circle.

Another problematic aspect of traversable wormholes is that
they have the potential to lead to causal inconsistencies. For example,
by applying a boost to one end of a wormhole one could attempt to
create a configuration with closed time-like curves \cite{Morris1988}. The direct interaction
of the boundaries that we require implies that no such paradoxes may arise (for a more detailed discussion, see section 4).

The traversable wormhole we find is the first such solution that has been shown to be embeddable in a UV complete theory of gravity. A phenomenological model of a static BTZ wormhole that becomes traversable as a result of nonperturbative effects in a $1/c$ expansion was proposed in \cite{Solodukhin:2005qy} ($c$ being the central charge), however it was not shown that the metric obeys any field equations. A traversable wormhole solution of five dimensional Einstein-Gauss-Bonnet gravity was found in \cite{Bhawal:1992sz, Thibeault:2005ha, Arias:2010xg}, however that low energy effective theory appears to lack a UV completion \cite{Camanho:2014apa}. Another example was found \cite{Barcelo:1999hq} in a theory with a conformally coupled scalar, in a regime in which the effective Newton's constant becomes negative. This suggests that this solution also cannot arise in a UV complete model. The important fact that the boundary CFT dual of a traversable wormhole must involve interactions between the two CFTs was noted in \cite{Solodukhin:2005qy, Arias:2010xg}.

\begin{figure}[h]
\begin{centering}
\subfloat[\label{fig:Penrose-diagram-of}]{\protect\begin{centering}
\protect\includegraphics[width=4cm]{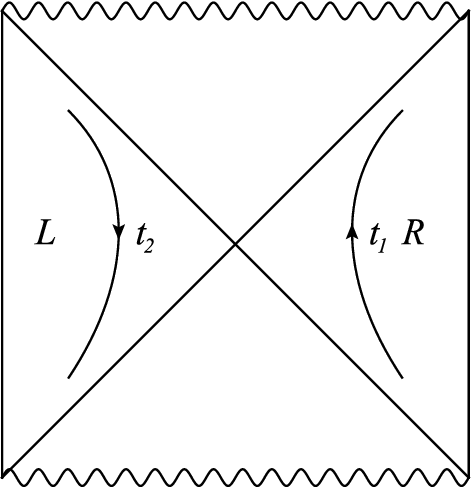}\protect
\par\end{centering}

}$\qquad$\subfloat[\label{fig:Kruskal}]{\protect\begin{centering}
\protect\includegraphics[width=4.5cm]{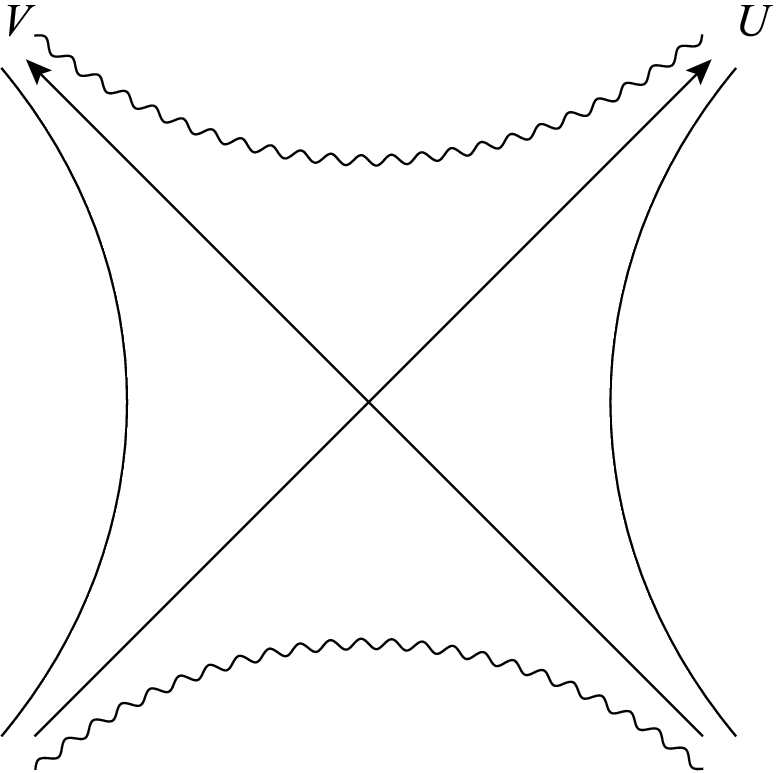}\protect
\par\end{centering}

}
\par\end{centering}

\protect\caption{(a) is the Penrose diagram and (b) shows the Kruskal coordinates of the eternal black
hole}
\end{figure}

The eternal black hole with two asymptotically AdS regions is the
simplest setting to investigate these questions \cite{Maldacena2003}.
We will deform the system by turning on a relevant double trace deformation \cite{Aharony2005}
\begin{equation}
\d S=\int dt\ d^{d-1}x\ h(t,x)\mO_{R}(t,x)\mO_{L}(-t,x), \label{eq:1.2}
\end{equation}
where ${\cal O}$ is a scalar operator of dimension less than $d/2$, dual to a scalar field $\varphi$.
This connects the boundaries with the same time orientation, since the $t$
coordinate runs in opposite directions in two wedges (see
Fig. \ref{fig:Penrose-diagram-of}). The small deformation $h(t,x)$ has support only after some turn-on time $t_{0}$. By the AdS/CFT
correspondence, we can be certain that this relevant deformation corresponds to a consistent
configuration in quantum gravity.

The eternal black hole has a Killing symmetry which is time-like outside
the horizon. Null rays along the horizon $V=0$ pass through the
bifurcation surface of the Killing vector, and asymptote to $t\rightarrow-\infty$
on the left boundary and $t\rightarrow+\infty$ on the right boundary
(see Fig. \ref{fig:Kruskal}). Denote the affine parameter along this ray as
$U$. In the linearized analysis around this solution, the throat
will become marginally traversable if $\int dU\ T_{UU}<0$, where the integral is over the whole $U$ coordinate.

It is instructive to see explicitly in this case that a small spherically symmetric perturbation of the stress tensor $T_{\mu\nu}\sim O(\e)$ results in a traversable wormhole exactly when the ANEC is violated, by solving the linearized Einstein equation for $h_{\mu\nu}=\d g_{\mu\nu}\sim O(\e)$.
Using Kruskal coordinates for the background metric, we find that at $V=0$,
\begin{equation}
\f{(d-2)}4\left[\left((d-3)r_{h}^{-2}+(d-1)\ell^{-2}\right)\left(h_{UU}+\del_{U}(Uh_{UU})\right)-2r_{h}^{-2}\del_{U}^{2}h_{\phi\phi}\right]= 8\pi G_N\, T_{UU}\label{eq:1.4}
\end{equation}
where $\phi$ is the
azimuthal angle, $r_{h}$ is the horizon
radius of the black hole and the cosmological constant is $\Lambda=-\f{(d-2)(d-1)}{2l^{2}}<0$.

Since the deformation of the Hamiltonian is small,
after the scrambling time, the fields ought to approach a stationary state with respect to an asymptotic Killing symmetry $U \partial U$.  Hence $T_{UU}$ must decay faster than $U^{-2}$, as does
each term in LHS of (\ref{eq:1.4}) after imposing a suitable gauge at past and future infinity.  Therefore, if we integrate (\ref{eq:1.4}) over $U$ the total derivative terms drop out and we obtain
\begin{equation}
8\pi G_N \int dU T_{UU}=\f{(d-2)}4\left((d-3)r_{h}^{-2}+(d-1)\ell^{-2}\right)\int dU h_{UU}
\end{equation}
Linearized diffeomorphisms around the stationary black hole background act on $h_{\mu \nu}$, but when the AdS asymptotic conditions are imposed the quantity $\int^{+\infty}_{-\infty} d U \ h_{U U}$ is gauge invariant.
Note that the null ray originating on the past horizon is given in coordinates by
\begin{equation}
V(U)=-(2g_{UV}(0))^{-1}\int_{-\infty}^{U}dU h_{UU}
\end{equation}
after including the perturbation to linear order,
where $g_{UV}(0)<0$ is the $UV$ component of the original metric on the $V=0$
slice. If the ANEC is violated, $V(+\infty)<0$, and a light
ray from left boundary will hit the right boundary after
finite time.

Note that if there existed any state in which the wormhole was traversable in
the system defined by the decoupled Hamiltonian, $H_{L}+H_{R}$, then
it would contradict the AdS/CFT duality. This is because in the decoupled
system, no operator on the left can influence the right, which implies
that no signal can be transmitted between the boundaries through the
bulk.

At the linearized level, if one modifies the state as $|\textrm{tfd}\rangle\rightarrow e^{i\epsilon A}|\textrm{tfd}\rangle$
for small $\epsilon$, the average null energy becomes $\langle\int dUT_{UU}\rangle=i\epsilon\langle[\int dUT_{UU},A]\rangle$.
If this were non-vanishing for any operator $A$, then by adjusting
the sign of $\epsilon$, the throat could be made traversable. It
is easy to check that the expectation value of this commutator indeed vanishes.

In fact, $\bra{\text{tfd}}$ is invariant under $H_{R}-H_{L}$, which
corresponds to the bulk Killing symmetry $i\del_{t}$ (note the
directions are opposite in left and right wedges). On the horizon $V=0$,
 one can show $\del_{t}=U\del_{U}$ in Kruskal coordinates, which
is just a dilation of the $U$ direction. Note that under the
$U\ra\lambda U$ scaling, $T_{UU}\ra\lambda^{-2}T_{UU}$ and $dU\ra\lambda dU$,
which implies $[H_{R}-H_{L},\int dUT_{UU}]=-i\int dUT_{UU}$. Therefore
\begin{equation}
(H_{R}-H_{L})\int dUT_{UU}|\textrm{tfd}\rangle=[H_{R}-H_{L},\int dUT_{UU}]|\textrm{tfd}\rangle+\int dUT_{UU}(H_{R}-H_{L})|\textrm{tfd}\rangle=-i\int dUT_{UU}|\textrm{tfd}\rangle.
\end{equation}

 This implies that $\int dUT_{UU}\bra{\text{tfd}}$ is either an eigenvector
of $H_{R}-H_{L}$ with eigenvalue $-i$, or identically zero. Since
$H_{R}-H_{L}$ is a Hermitian operator, whose eigenvalues must be real,
it follows that $\int dUT_{UU}|\textrm{tfd}\rangle=0$. In
other words, $T_{UU}$ in the modified state along $U>0$ will exactly
cancel that along $U<0$. Beyond the linearized level, one can show that the backreaction
always causes the throat to lengthen \cite{Maldacena2013, Shenker2013}, so that
it cannot be traversed in any state of the decoupled system, as expected.

We will consider a deformation of the Hamiltonian that turns on at
some time $t_{0}$ in (\ref{eq:1.2}).\footnote{We do not consider the case of a time-independent interaction, in order to prevent the quantum state from becoming non-regular on the past horizon.}  At the linearized level, this
has the same effect as changing the state to the future of $t_{0}$.
Now there is no reason for the above cancellation to occur since $T_{UU}$
along $U<0$ is unchanged. Therefore, one expects that generically
by an appropriate choice of sign one will render the Einstein-Rosen bridge traversable, as long as the deformation couples the two boundaries.\footnote{A deformation of only $H_R$  has the same effect on the ANE as a change in the state, by bulk causality, since the past causal cone of the deformation does not intersect the $V=0$ null sheet. This again agrees with the fact that when the boundaries are decoupled, no traversable wormhole can exist.}

The simplest option in the large $N$ limit is a double trace deformation. This has the effect of modifying the boundary conditions for the dual scalar field, such that some amplitude of a wave hitting one boundary will be transmitted to the opposite one. This does not change the eternal black hole solution classically, but results in a quantum correction to the matter stress tensor.

In order to be sure that the configuration is an allowed one, we choose the deformation to be relevant. Then it will be a renormalizable deformation of the CFT, and the dual geometry will not be modified by backreaction in an uncontrolled way at the AdS boundaries. Also, heuristically, the effect of such a deformation coupling the two CFT's should be strong in the IR, which suggests that it renders the deep interior traversable.

Recall that the conformal weight of a scalar operator $\mO_{i}$
is given by $\Delta=\f d2\pm\sqrt{(\f d2)^{2}+M^{2}}$, where $M$ is the mass of the bulk field, and the plus
or minus sign depends on the choice of asymptotic boundary conditions.
In the case $M^{2}>0$, only the plus sign leads to normalizable
modes. However, unitarity in AdS space \cite{Breitenlohner1982} allows
a slightly tachyonic bulk field: $M^{2}>-(\f d2)^{2}$, in which
modes of both signs are normalizable and we are free to choose either
one. To have a relevant deformation, we start with the alternative boundary condition, associated with the minus sign.


A brief overview of this paper is as follows. In section 2, we calculate the bulk two-point function with the modified Hamiltonian at linear order in $h$. In section 3, we use the point-splitting method to calculate $T_{UU}$ on the $V=0$ slice. Numerical result shows that $T_{UU}$ is rendered negative by our boundary interaction. We find an analytic expression for $\int dU T_{UU}$, which is negative for all $0<\Delta<1$. In section 4 we calculate the energy and entropy of the resulting CFT state, and describe their holographic bulk duals.  In section 5, we discuss the properties of this traversable wormhole and propose a quantum teleportation interpretation in the ER=EPR context. The appendix is a detailed calculation of $\int dU T_{UU}$.

Throughout we use units where $c = \hbar = 1$.

\section{Modified bulk two-point function}

For simplicity, we consider the eternal BTZ  black hole \cite{Banados1992,Banados1993}
(for a review, see \cite{Carlip1995}), whose metric is
\begin{equation}
ds^{2}=-\f{r^{2}-r_{h}^{2}}{\ell^{2}}dt^{2}+\f{\ell^{2}}{r^{2}-r_{h}^{2}}dr^{2}+r^{2}d\phi^{2}\label{eq:2.1-1}
\end{equation}
The inverse temperature of the BTZ black hole is determined by its horizon
radius $r_{h}$ as $\b=2\pi\ell^{2}/r_{h}$. Here and below we set
AdS length $\ell$ to 1. Without any deformation of the Hamiltonian, the
bulk free field two-point function in the BTZ background with $r^{-\D}$
fall-off was first derived by the mode sum method in \cite{Ichinose1995}.

In right wedge, it is
\begin{equation}
\avg{\varphi_{R}(x)\varphi_{R}(x')}_{0}=\frac{1}{2^{3-\Delta}\pi}(G_{+}+G_{-})(G_{+}^{-1}+G_{-}^{-1})^{1-2\Delta}\label{eq:2.1}
\end{equation}
where
\begin{equation}
G_{\pm}\equiv\left(\frac{rr'}{r_{h}^{2}}\cosh r_{h}\Delta\phi\pm1-\frac{(r^{2}-r_{h}^{2})^{1/2}(r'^{2}-r_{h}^{2})^{1/2}}{r_{h}^{2}}\cosh r_{h}\Delta t\right)^{-1/2}.
\end{equation}
The bulk field operator $\varphi_{R}(x)$ in the eternal black hole background
can be understood as a non-local CFT operator \cite{Papadodimas:2012aq}. In particular, $\varphi_{R}(x)$
can be expanded in terms of the right boundary dual operator as
\begin{equation}
\varphi_{R}(t,r,\phi)=\int_{\w>0}d\w\,dm\left(f_{\w m}(r)e^{-i\w t+im\phi}\mO_{\w m}+f_{\w m}^{*}(r)e^{i\w t-im\phi}\mO_{\w m}^{\dagger}\right)
\end{equation}
where $f_{\w m}(r)e^{-i\w t+im\phi}$ are bulk positive frequency normalizable
modes approaching $r^{-\D}$ when $r\ra\infty$ and $\mO_{\w m}$ is
the boundary annihilation operator defined by
\begin{equation}
\mO(t,\phi)=\int d\w\,dm\left(e^{-i\w t+im\phi}\mO_{\w m}+e^{i\w t-im\phi}\mO_{\w m}^{\dagger}\right).
\end{equation}

Therefore, the bulk to boundary correlation function is given by
\begin{align}
K_{\D}(r,t,\phi) & \equiv\avg{\varphi_{R}(t,r,\phi)\mO(0,0)}=\lim_{r'\ra\infty}r'^{\D}\avg{\varphi_{R}(t,r,\phi)\varphi_{R}(0,r',0)}_{0}\nonumber \\
 & =\f{r_{h}^{\D}}{2^{\D+1}\pi}\left(-\f{(r^{2}-r_{h}^{2})^{1/2}}{r_{h}}\cosh r_{h}t+\f r{r_{h}}\cosh r_{h}\phi\right)^{-\D},\label{eq:2.5}
\end{align}
where we used translation symmetry in $t$ and $\phi$
to move $(t',\phi')$ to the boundary origin. This expression is real
only when $(r,t,\phi)$ is space-like separated from the boundary origin.
For time-like separation, general analytic properties of Wightman functions imply that
one should change $t$ to $t-i\e$, which assigns a phase of $e^{-i\pi\Delta}$
when $t>0$ and of $e^{i\pi\D}$ when $t<0$.

Now we consider the time dependent modified Hamiltonian of (\ref{eq:1.2}):
\begin{equation}
\d H(t)=-\int d\phi\,h(t,\phi)\mO_{R}(t,\phi)\mO_{L}(-t,\phi),\label{eq:2.6}
\end{equation}
where 
$h(t,\phi)=0$ when $t<t_{0}$. Using evolution operator $U(t,t_{0})=\mT e^{-i\int_{t_{0}}^{t}dt\d H(t)}$ in interaction
picture, the bulk two-point function is
\begin{equation}
\avg{\varphi^H_{R}(t,r,\phi)\varphi^H_{R}(t',r',\phi')}=\avg{U^{-1}(t,t_{0})\varphi^I_{R}(t,r,\phi)U(t,t_{0})U^{-1}(t',t_{0})\varphi^I_{R}(t',r,\phi)U(t',t_{0})}\label{eq:2.7}
\end{equation}
where superscripts $H$ and $I$ represent Heisenberg and interaction picture respectively. To leading order in $h$, (\ref{eq:2.7})
is (suppressing $r$ and $\phi$ coordinates and omitting $I$)
\begin{align}
G_{h}\equiv & -i\int_{t_{0}}^{t}dt_{1}h(t_{1})\avg{[\mO_{L}(-t_{1})\mO_{R}(t_{1}),\varphi_{R}(t)]\varphi_{R}(t')}-i\int_{t_{0}}^{t'}dt_{1}h(t_{1})\avg{\varphi_{R}(t)[\mO_{L}(-t_{1})\mO_{R}(t_{1}),\varphi_{R}(t')]}\nonumber \\
\simeq & -i\int_{t_{0}}^{t}dt_{1}h(t_{1})\avg{\varphi_{R}(t')\mO_{L}(-t_{1})}\avg{[\mO_{R}(t_{1}),\varphi_{R}(t)]}+(t\leftrightarrow t')\nonumber \\
= & i\int_{t_{0}}^{t}dt_{1}h(t_{1})K_{\D}(t'+t_{1}-i\b/2)\left[K_{\D}(t-t_{1}-i\e)-K_{\D}(t-t_{1}+i\e)\right]+(t\leftrightarrow t')\nonumber \\
= & 2\sin\pi\D\int dt_{1}h(t_{1})K_{\D}(t'+t_{1}-i\b/2)K_{\D}^{r}(t-t_{1})+(t\leftrightarrow t')\label{eq:2.8}
\end{align}
where in the second line we used large $N$ factorization and causality,
in that $\mO_{L}$ commutes with any $\varphi_{R}$, in the third line
we used the KMS condition \cite{Haag1967}
\begin{equation}
\avg{\mO_{R}(t)\mO_{L}(t')}_{tfd}=\avg{\mO_{R}(t)\mO_{R}(t'+i\b/2)}_{tfd}
\end{equation}
and in the last line $K_{\D}^{r}$ is the retarded correlation function
\begin{equation}
K_{\D}^{r}(t,r,\phi)=|K_{\D}(t,r,\phi)|\ \theta(t)\ \theta\left(\f{(r^{2}-r_{h}^{2})^{1/2}}{r_{h}}\cosh r_{h}t-\f r{r_{h}}\cosh r_{h}\phi\right)\label{eq:2.10}
\end{equation}

One can also derive (\ref{eq:2.8}) using the bulk mode sum method with
modified boundary conditions. This approach would allow one to compute the stress tensor for finite $h$, not just perturbatively. According to Lorentzian AdS/CFT, the
double trace deformation \cite{Witten2001,Berkooz2002}, from  the point
of view of the right wedge, is equivalent to a source term $h(t,\phi)\mathcal{O}_{L}(-t,\phi)$,
for $\mathcal{O}_{R}(t)$,
activating the initially frozen fall-off component of the bulk field.
The same applies to the left
wedge. Therefore the asymptotic behavior of a global bulk mode
$\varphi$ living in the entire eternal black hole should satisfy
\begin{align}
\varphi(r\ra\infty_{R}) & \ra\a_{R}(t,\phi)r^{-\D}+\b_{R}(t,\phi)r^{-2+\D},\ \b_{L}(t,\phi)=h(-t,\phi)\a_{R}(-t,\phi)\label{eq:2.11}\\
\varphi(r\ra\infty_{L}) & \ra\a_{L}(t,\phi)r^{-\D}+\b_{L}(t,\phi)r^{-2+\D},\ \b_{R}(t,\phi)=h(t,\phi)\a_{L}(-t,\phi)\label{eq:2.12}
\end{align}
where the subscript $1$ is for right wedge and $2$ is for left wedge.

The thermofield double state of the eternal black hole is the vacuum state
in the Kruskal patch \cite{Israel1976}. This is analogous to the relation
between the Minkowski vacuum and the Rindler thermofield double state \cite{Unruh1976}.
Choosing the appropriate global bulk modes $H_{\w m}^{(\pm)}$\footnote{This step is very tricky because at order $h$, the  $r^{-\D}$
component is not constrained by the deformation. The only requirement
 is that the modified two point function must be regular on
horizon. We were able to find a choice to reproduce (\ref{eq:2.8})
up to normalization. } and applying the method of \cite{Israel1976}, we can construct $\varphi$ as
\begin{equation}
\varphi(x)=\int_{\w>0}d\w\,dm(H_{\omega m}^{(+)}(x)b_{\omega m}^{(+)}+H_{\omega m}^{(-)}(x)b_{\omega m}^{(-)\dagger}+h.c.)
\end{equation}
where $b_{\w m}^{(\pm)}$ are annihilation operators used to define the
vacuum. We find the two-point function in this vacuum is the same
as (\ref{eq:2.8}) up to normalization. Since the calculation is quite
involved, we do not include it in this paper.

\section{1-loop stress tensor}

The stress tensor is given by variation of action with respect to
$g^{\mu\nu}$,
\begin{equation}
T_{\mu\nu}=\partial_{\mu}\varphi\partial_{\nu}\varphi-\frac{1}{2}g_{\mu\nu}g^{\rho\sigma}\partial_{\rho}\varphi\partial_{\sigma}\varphi-\frac{1}{2}g_{\mu\nu}M^{2}\varphi^{2}
\end{equation}
The 1-loop expectation value can be calculated by point splitting,
\begin{align}
\langle T_{\mu\nu}\rangle & =\lim_{x\rightarrow x'}\partial_{\mu}\partial'_{\nu}G(x,x')-\frac{1}{2}g_{\mu\nu}g^{\rho\sigma}\partial_{\rho}\partial'_{\sigma}G(x,x')-\frac{1}{2}g_{\mu\nu}M^{2}G(x,x')\label{eq:3.2}
\end{align}
where $G(x,x')$ is 2-point function. In this formula, one must renormalize the stress tensor by subtracting the coincident point singularities from the 2-point function, which are given by the Hadamard conditions \cite{parker2009quantum}.
Since these are determined by the short distance dynamics, this subtraction is unchanged when we modify the boundary conditions, and it has no effect on the order $h$ correction that we are interested in.

At leading order, as we reviewed in the Introduction, $\int dUT_{UU}$ is zero on the horizon $V=0$. Indeed, the leading order two point function
in the BTZ black hole is (\ref{eq:2.1}) where $\phi$ has periodicity  $2\pi$
and all $\D\phi+2\pi n$ images are summed. The only coincident point pole
comes from the $n=0$ component. Summing over the other $n$ components, one finds  that in Kruskal coordinates the leading order stress
tensor $T_{UU}\sim O(V^{2})$ in the $V\ra0$ limit, so that $T_{UU}=0$
along the horizon.

The subleading 2-point function is given by (\ref{eq:2.8}). Note
that $h(t,\phi)$ is dimensionful and its dimension is $2-2\D$ because
in (\ref{eq:2.6}) $\mO$ has dimension $\D$\footnote{Here we implicitly defined the unit length angular coordinate $x\equiv\phi\ell$.
Taking the limit $r\ra\infty$ in BTZ metric (\ref{eq:2.1-1}), the boundary
metric is flat $ds_{b}^{2}=-dt^{2}+dx^{2}$.}. Moreover, since $h(t,\phi)$ is a boundary CFT smearing function, it
should not depend on any bulk length scale (e.g. $r_{h}$ and $\ell$)
explicitly but only on the inverse temperature $\b$. Let us assume that
$h(t,\phi)$ is uniform over $\phi$:
\begin{equation}
h(t,\phi)=\begin{cases}
h(2\pi/\b)^{2-2\D} & t\geq t_{0}\\
0 & t<t_{0} \label{eq:3.3}
\end{cases}
\end{equation}
where $h$ is a dimensionless constant. In Kruskal coordinates
\begin{equation}
e^{2r_{h}t}=-\frac{U}{V},\;\frac{r}{r_{h}}=\frac{1-UV}{1+UV}
\end{equation}
the change in the 2-point function  is
\begin{align}
G_{h}= & C_{0}\left(\f{2\pi}{\b}\right)^{2\D-2}r_{h}\int\f{dU_{1}}{U_{1}}d\phi_{1}h(U_{1},\phi_{1})\left(\f{1+UV}{U/U_{1}-VU_{1}-(1-UV)\cosh r_{h}(\phi-\phi_{1})}\right)^{\D}\nonumber \\
 & \times\left(\f{1+U'V'}{U'U_{1}-V'/U_{1}+(1-U'V')\cosh r_{h}(\phi'-\phi_{1})}\right)^{\D}+(U,V,\phi\leftrightarrow U'V',\phi')
\end{align}
where $C_{0}=\f{r_{h}^{2\D-2}\sin\D\pi}{2(2^{\D}\pi)^{2}}\left(\f{2\pi}{\b}\right)^{2-2\D}$
and we transformed the integral over $t_{1}$ to Kruskal coordinates in which the boundary is $U_{1}V_{1}=-1$. Note that this result applies to both the black hole and black brane cases because the integration
of $\phi_{1}$ over $0$ to $2\pi$ and summing over $n$ with modification
$\phi_{1}\ra\phi+2\pi n$ is equivalent to the integration of $\phi_{1}$
over the whole real axis. Since we only focus on $T_{UU}$ component on the
horizon $V=0$ and the derivative on $U$ and $U'$ in (\ref{eq:3.2})
has nothing to do with the value of $V$ and $V'$, we can take both
points to the horizon first, namely $V=V'=0$. Similarly, we can take
$\phi=\phi'$ first for simplicity. Since $h(t_{1},\phi_{1})$ is
uniform in $\phi_{1}$, $\del_{\phi}$ is still a Killing vector of
the system and therefore $G_{h}$ should not depends on $\phi$. Defining
$y=\cosh r_{h}(\phi_{1}-\phi)$, on horizon we have
\begin{equation}
G_{h}=hC_{0}\int_{U_{0}}^{U}\f{dU_{1}}{U_{1}}\int_{1}^{\f U{U_{1}}}\f{2dy}{\sqrt{y^{2}-1}}\left(\f{U_{1}}{U-U_{1}y}\right)^{\D}\left(\f 1{U'U_{1}+y}\right)^{\D}+(U\leftrightarrow U')\equiv F(U,U')+F(U',U)\label{eq:3.6}
\end{equation}
where $U_{0}=e^{r_{h}t_{0}}$. The integral range of (\ref{eq:3.6})
is given by the step function in (\ref{eq:2.10}), which ensures that $U-U_{1}y\geq0$.
Note that the integral in (\ref{eq:3.6}) is dimensionless. Since
$G_{h}$ has dimension 1 ($\varphi_{R}$ has dimension $\f 12$ in 3-dimension
spacetime), if we restore $\ell$ in (\ref{eq:3.6}), we find the
total length scale dependence of $G_{h}$ is $\ell^{-1}$.

\begin{figure}
\begin{centering}
\subfloat[\label{fig:3.1a}]{\protect\begin{centering}
\protect\includegraphics[width=8cm]{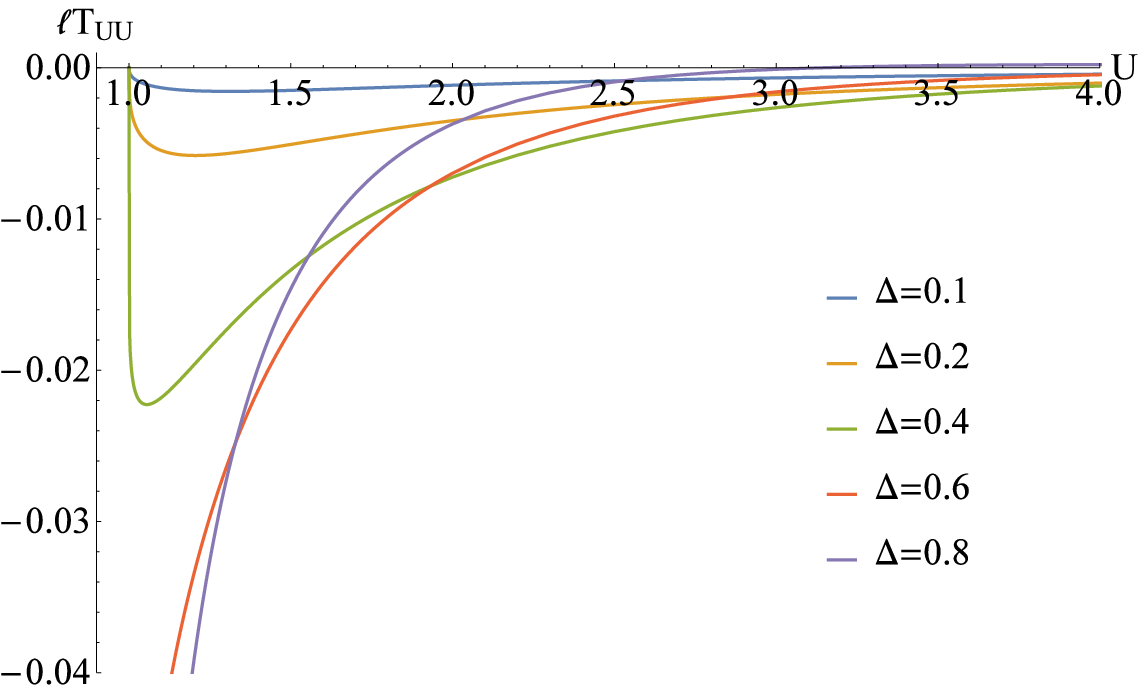}\protect
\par\end{centering}

}\subfloat[\label{fig:3.1b}]{\protect\begin{centering}
\protect\includegraphics[width=8cm]{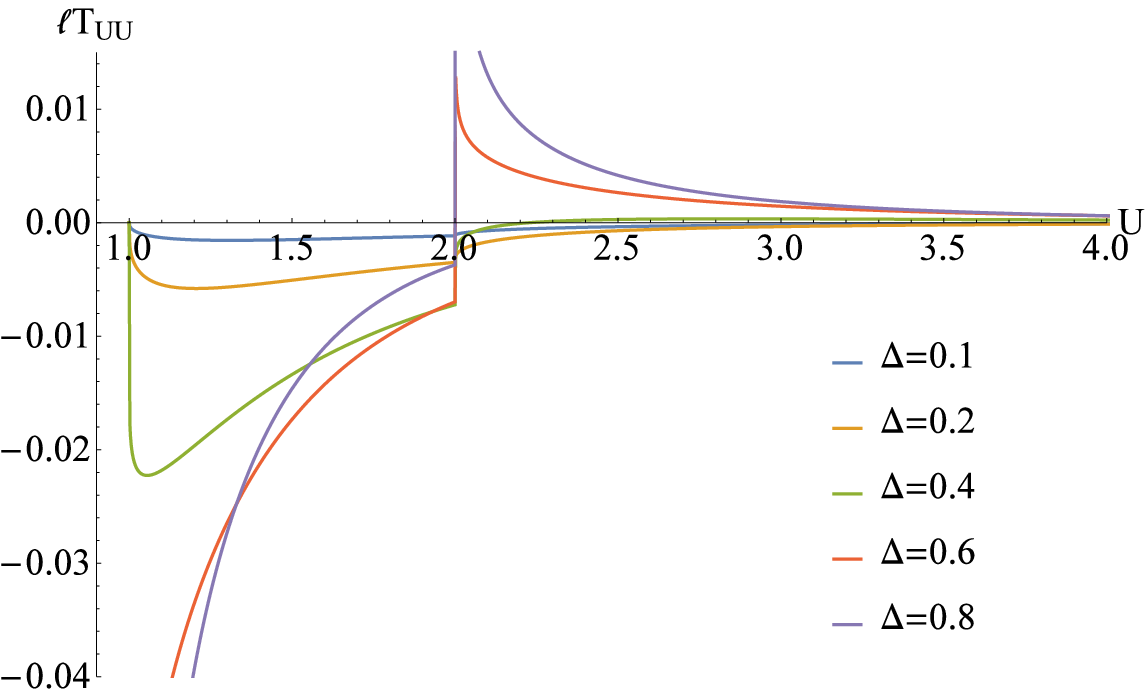}\protect
\par\end{centering}

}
\par\end{centering}

\protect\caption{(a) shows the null energy along the horizon when the interaction is turned on at $U=U_{0}=1$ and never
shut off, with our choice for the sign of the coupling $h$; (b) shows the case where it is turned on at $U=U_{0}=1$ and turned
off at $U=U_{f}=2$. In both cases, $h=1$. We see clearly in both
(a) and (b) that $T_{UU}$ becomes negative after turn-on; in (b) $T_{UU}$
becomes positive after turn-off. Blue is for $\protect\D=0.1$; yellow
is for $\protect\D=0.2$; green is for $\protect\D=0.4$; pink is
for $\protect\D=0.6$; purple is for $\protect\D=0.8$}
\end{figure}

Note that $g_{UU}=0$ in the original BTZ geometry. By (\ref{eq:3.2}),
$T_{UU}$ on horizon is
\begin{equation}
T_{UU}=\lim_{U'\ra U}\del_{U}\del_{U'}(F(U,U')+F(U',U))=2\lim_{U'\ra U}\del_{U}\del_{U'}F(U,U')
\end{equation}
where we should note the dimension of $T_{UU}$ is the same as $G_{h}$
because $U$ is dimensionless. Since the integration ranges are only
functions of $U$, we can take the $U'$ derivative before evaluating
the integral
\begin{equation}
T_{UU}=-4h\D C_{0}\lim_{U'\ra U}\del_{U}\int_{U_{0}}^{U}dU_{1}\int_{1}^{\f U{U_{1}}}\f{dy}{\sqrt{y^{2}-1}}\f{U_{1}^{\D}}{(U-U_{1}y)^{\D}(U'U_{1}+y)^{\D+1}}\label{eq:3.8}
\end{equation}
Defining a new variable $x=\f{y-1}{U/U_{1}-1}$ and
integrating over $x$ we get
\begin{align}
T_{UU}= & -\f{4h\D C_{0}\G(\f 12)\G(1-\D)}{\sqrt{2}\Gamma(\f 32-\D)}\lim_{U'\ra U}\del_{U}\int_{U_{0}}^{U}dU_{1}\f{F_{1}(\f 12;\f 12,\D+1;\f 32-\D;\f{U_{1}-U}{2U_{1}},\f{U_{1}-U}{U_{1}(1+U'U_{1})})}{U_{1}^{-\D+1/2}(U-U_{1})^{\D-1/2}(1+U'U_{1})^{\D+1}}\label{eq:3.9}
\end{align}
where we used the integral representation of Appell hypergeometric
function. The integral over $U_{1}$ is finite as long as $\D-1/2<1$,
namely $\D<3/2$, because in the integrated region, the only potentially
divergent point is around $U_{1}\ra U$ from below since $F_{1}$
is a complete function when $\D<3/2$. In particular, when $U_{1}\sim U$,
$F_{1}\ra1$, which implies $\D<3/2$ is the sufficient and necessary
condition for integrability. Defining a new variable $z=\f{U_{1}-U_{0}}{U-U_{0}}$,
the domain of integration in (\ref{eq:3.9}) becomes 0 to 1 and therefore
we can exchange the order of $\del_{U}$ and $\int dz$. After differentiating
w.r.t. $U$, and restoring the variable $U_{1}$, we get
\begin{align}
T_{UU}= & -\f{2h\D C_{0}\G(\f 12)\G(1-\D)}{\Gamma(\f 32-\D)}\int_{U_{0}}^{U}\f{dU_{1}U_{1}^{2\D}(f_{1}+f_{2}+f_{3})}{(U-U_{0})(U-U_{1})^{\D-1/2}(1+U_{1}^{2})^{\D+1}U^{\D+1}(U+U_{1})^{1/2}}\label{eq:3.10}
\end{align}
where
\begin{align}
f_{1} & =\f{-2\Delta(UU_{1}^{2}+U_{0})+3UU_{0}U_{1}+U_{0}+2U_{1}}{1+UU_{1}}F_{1}(1-\D,\f 12,1+\D,\f 32-\D,u,v)\\
f_{2} & =\f{2(1+\D)(U-U_{1})(U_{0}+2UU_{0}U_{1}-UU_{1}^{2})}{(2\D-3)U(1+U_{1}^{2})(1+UU_{1})}F_{1}(1-\D,\f 12,2+\D,\f 52-\D,u,v)\\
f_{3} & =\f{U_{0}(U-U_{1})}{(2\D-3)(U+U_{1})}F_{1}(1-\D,\f 32,1+\D,\f 52-\D,u,v)\\
u & =\f{U-U_{1}}{U+U_{1}},\qquad v=\f{U-U_{1}}{U(1+U_{1}^{2})}
\end{align}
Performing the final integral numerically, we plot the result in
Fig. \ref{fig:3.1a}.

In the figure, we see that the null energy is negative after we turn on the insertion
at $U_{0}=1$ if we take positive $h$. Physically,
this means the light-like ray $V=0$ becomes time-like after $U_{0}$
and a spaceship that enters early enough may  escape the black hole!

One may note that when $\D<1/2$, $T_{UU}$ is finite but when $\D>1/2$,
$T_{UU}$ is singular near insertion time $U_{0}$. However, this
singularity is not essential because it is integrable, as we will see
later when we calculate $\int dUT_{UU}$ along the horizon $V=0$. Indeed,
the classical solution of Einstein equations for a shockwave insertion
in the bulk in Kruskal coordinates contains a delta function, which
is also an integrable singularity \cite{Shenker2013}. One might also worry that the derivative of $g_{UU}$ and the Riemann curvature are singular
at the turn-on and turn-off times,
although $T_{UU}$ and $\int dUT_{UU}$ are not. However, this is simply due
to the fact that we turned the insertion on and off  as a step function.
If this process were taken to be smooth enough, there would be no singularity.

To see the late time behavior, we can use the $z$ variable to rewrite
(\ref{eq:3.10}) in the large $U$ limit. In this limit, $f_{1}$ dominates among all $f_{i}$'s
in (\ref{eq:3.10}). Using the identity $F_{1}(a;b,b';c;z,0)={}_{2}F_{1}(a,b;c;z)$
we obtain
\begin{equation}
T_{UU}\ra\f{4h\D^{2}C_{0}\G(\f 12)\G(1-\D)}{\Gamma(\f 32-\D)U^{2\D+2}}\int_{0}^{1}\f{dz\,z^{2\D+1}{}_{2}F_{1}(1-\D,\f 12,\f 32-\D,\f{1-z}{1+z})}{((z+\e)^2+\e)^{\D+1}(1-z)^{\D-1/2}(1+z)^{1/2}}\ra0_{+}
\end{equation}
where $\e$ is a small number of order $U^{-1}$ and which implies that $T_{UU}$ becomes positive and decays to zero at
late times.

If we turn off the interaction at some finite time $U_{f}$, when $U>U_f$, we can safely pass $\del_{U}$ into the $U_{1}$ integral,
which leads to
\begin{align}
T_{UU} & =-\f{4h\D C_{0}\G(\f 12)\G(1-\D)}{\Gamma(\f 12-\D)}\int_{U_{0}}^{U_{f}}dU_{1}\f{U_{1}^{2\D+1}F_{1}(-\D;\f 12,\D+1;\f 12-\D;\f{U-U_{1}}{U+U_{1}},\f{U-U_{1}}{U(1+U_{1}^{2})})}{(U-U_{1})^{\D+1/2}(U+U_{1})^{1/2}U^{\D+1}(1+U_{1}^{2})^{\D+1}}\label{eq:3.16}
\end{align}
In deriving (\ref{eq:3.16}), we used a property of the derivative of the
Appell hypergeometric function and equation (7a) in \cite{Schlosser2013}.
The numerical result is plotted in Fig. \ref{fig:3.1b}.

In this figure, we see that after turning off the interaction, $T_{UU}$
has a jump and becomes positive at late times. In particular, when
$\D>1/2$, $T_{UU}$ becomes divergent again right after $U_{f}$.
Fortunately, it is again an integrable divergence which should not
cause any physical problem. By the identity \cite{Prudnikov1992}:
\begin{equation}
F_{1}(a;b,b';c;x,y)=\sum_{m\geq0}\f{(a)_{m}(b)_{m}}{m!(c)_{m}}x^{m}{}_{2}F_{1}(a+m,b';c+m;y)\label{eq:3.17-1}
\end{equation}
the late time behavior can be analyzed by taking the $U\ra\infty$ limit
in (\ref{eq:3.16}):
\begin{align*}
T_{UU} & \sim\f{4h\D^{2}C_{0}}{U^{2\D+2}}\log U\log\f{U_{f}}{U_{0}}\ra0_{+}
\end{align*}
Again, we find $T_{UU}$ becomes positive after some time and decays
to zero. In both late time analyses, $T_{UU}$ decays like
$U^{-2\D-2}$, which validates the assumption that $Uh_{UU}$ and
$\del_{U}h_{\phi\phi}$ vanish when $U\ra\infty$ in (\ref{eq:1.4}).

\begin{figure}
\begin{centering}
\includegraphics[width=10cm]{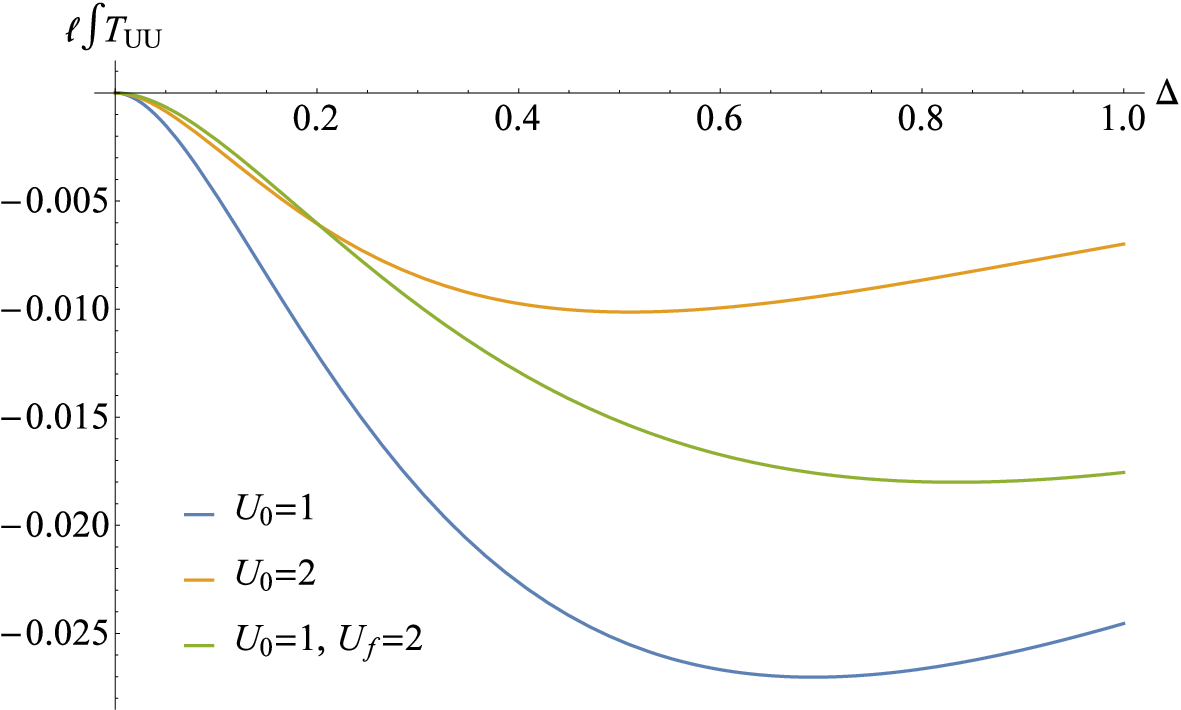}
\par\end{centering}

\protect\caption{$\int dUT_{UU}$ as a function of $\protect\D$; blue is for $U_{0}=1$;
yellow is for $U_{0}=2$; green is for $U_{0}=1$ and $U_{f}=2$\label{fig:intTUU}}
\end{figure}

In the above discussion, we see that at some finite time $T_{UU}$ becomes
positive whether or not we turn off the insertion, which might appear dangerous for the fate of the worm hole. The crucial diagnostic is
the sign of the integral of $T_{UU}$ over the
whole $V=0$ slice. This is what determines whether a light ray on horizon
eventually reaches the boundary at spatial infinity.

It looks horrible to integrate
$U$ in (\ref{eq:3.10}) from $U_{0}$ to infinity. Interestingly
and surprisingly, by some tricks, we can get a closed form for
it (see Appendix \ref{sec:intTUU}):
\begin{equation}
\int_{U_{0}}^{\infty}dUT_{UU}=-\f{h\G(2\D+1)^{2}}{2^{4\D}(2\D+1)\Gamma(\D)^{2}\G(\D+1)^{2}\ell}\f{_{2}F_{1}(\f 12+\D,\f 12-\D;\f 32+\D;\f 1{1+U_{0}^{2}})}{(1+U_{0}^{2})^{\D+1/2}}\label{eq:3.17}
\end{equation}
If we turn off the interaction at $U_{f}$, the integral is just the
difference between $\int_{U_{0}}^{\infty}dUT_{UU}$ and $\int_{U_{f}}^{\infty}dUT_{UU}$.
We plot the result as a function of $\D$ in Fig. \ref{fig:intTUU}.

In this figure, we see that for all $\D$ values from $0$
to $1$, the integral of $T_{UU}$ is always negative, which demonstrates the existence of a traversable wormhole.
Furthermore, the earlier
we turn on the insertion, the larger the effect is. In particular,
even if $T_{UU}$ becomes positive in late times, the wormhole still
exists since the integral of $T_{UU}$ remains negative.
Note that $\D=0$ is a special case where $\int dUT_{UU}=0$. Indeed,
the only $\D=0$ operator in CFT is the identity and of course adding
the product of identity operators to Hamiltonian has no effect on the
system.

\section{Holographic Energy and Entropy}\label{HEE}


In this section we will consider the implications of a traversable wormhole for the holographic entanglement entropy conjecture, which in this context relates the entanglement entropy between the two boundary CFT's to the area/entropy of certain extremal surfaces in the bulk theory \cite{Ryu:2006bv,Hubeny:2007xt,Barrella:2013wja,Faulkner:2013ana,EngelhardtWall}.

As a preliminary, we discuss the change of energy of the CFT state.  Long after the interaction is shut off, the system returns to thermal equilibrium.  Thus the final horizon area can be determined from the energy of the system, measured on the left or the right.  It is straightforward to check that, in our state, the energy decreases at linear order in $h$ with the sign choice that rendered the wormhole traversable:

After deforming the Hamiltonian $(t>t_{0})$, the state in Schr\"{o}dinger
picture is
\begin{equation}
\bra{\Psi(t)}=e^{-iH_{0}(t-t_{0})}U(t,t_{0})\bra{\text{tdf}}.
\end{equation}
Expanding $U(t,t_{0})$ to leading order in $h(t)$ given by (\ref{eq:3.3}), we find that the change in the energy on the right is
\begin{align}
\d E_R & =
i\int_{t_{0}}^{t}dt_{1}h(t_{1})\ket{\text{tdf}}[\d H(t_{1}),H_{R}]\bra{\text{tdf}}\nonumber \\
 & =\int_{t_{0}}^{t}dt_{1}d\phi h(t_{1})\ket{\text{tdf}}\del_{t}\mO_{R}(t_{1},\phi)\mO_{L}(-t_{1},\phi)\bra{\text{tdf}}\nonumber \\
 & =\f{hr_h^2}{2^{\D+1}\ell^3}\sum_{n}\left(\f 1{(\cosh2r_{h}t+\cosh2\pi r_{h}n)^{\D}}-\f 1{(\cosh2r_{h}t_{0}+\cosh2\pi r_{h}n)^{\D}}\right)\label{eq:12}
\end{align}
where in the second line we used the Heisenberg equation and in last line the boundary two-point function is obtained by taking limit  $r\ra\infty$
in (\ref{eq:2.1}) where $\phi$ has period $2\pi$, and all of its
images are summed over in the global BTZ black hole.\footnote{We consider global AdS here so that the total energy is finite.} If the interaction shuts off at $t_{f}$, the energy obviously becomes constant for $t>t_f$, and
$t$ in (\ref{eq:12}) is replaced by $t_{f}$.
Therefore, the effect of the interaction with $h>0$ is to reduce the energy.  Note that if there are any UV divergences in the energy they cannot appear at linear order in $h$, since the interaction involves just one field in each CFT.

At least at first order in $h$, the entropy of entanglement $S_{EE}$ between the left and right boundaries should also be well-defined (and time dependent) even during the period of time when the interaction is turned on, if one thinks of the state as evolving by the deformed Hamiltonian in the original tensor product Hilbert space.
By the first law of entanglement, at linear order in $h$, the change in $S_{EE}$ is equal to $\beta \delta H_R$, thus it also decreases until the turn-off time $t_f$ after which it remains constant (as it must under decoupled unitary evolution on the left and right).

The change in $S_{EE}$ is ${\cal O}(1)$ in a $1/c \sim 1/N \sim G_N/L_{AdS}$ expansion.\footnote{These are the correct scaling relations for $2+1$ dimensional bulk. In other dimensions, the scaling with $G_N$ is the same, but the scaling with the number of species $N$ may vary.}  At this order, in the bulk interpretation $S_{EE}$ has two parts, namely the small gravitational correction to the area $A/4G_N$ of the extremal surface, and the entanglement entropy of bulk fields $S_\text{bulk}$ on the spacelike slice from the extremal surface to the boundary slice at time $t$ \cite{Faulkner:2013ana,Barrella:2013wja}. 

In our situation, causality implies that the geometry near the bifurcation surface is unaffected by the perturbation. Thus, at order $h$, the area of the quantum extremal surface is unchanged from the original state.  On the other hand, $S_\text{bulk}$ has nonlocal aspects.  Therefore, the decrease of $S_{EE}$ at first order must be entirely due to a corresponding decrease in $S_\text{bulk}$ evaluated at the bifurcation surface.

\begin{figure}[ht]
\begin{centering}
\includegraphics[width=8cm]{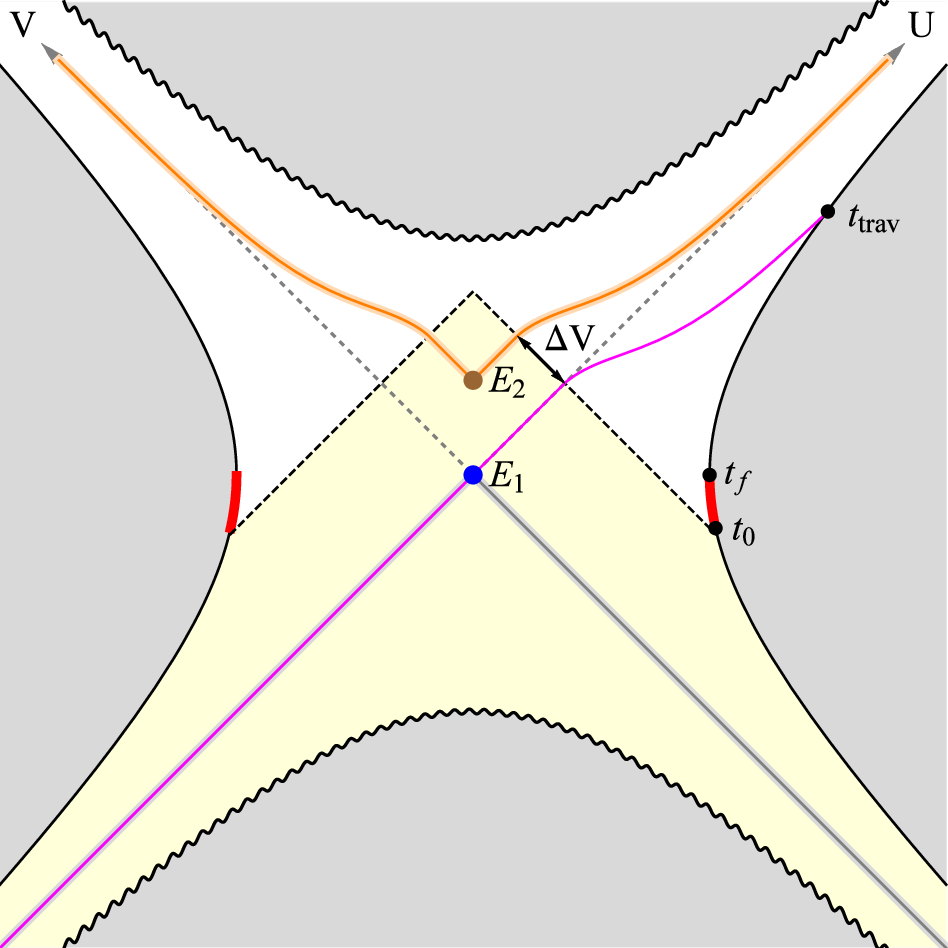}
\par\end{centering}

\protect\caption{The throat size is $\D V\sim h$. The red thick interval on the boundary is the duration of the deformation beginning at $t_0$ and ending at $t_f$. The metric in the light yellow region is unchanged and only that of the white region will have a nonzero backreaction correction. The orange thick curve is the future event horizon and the grey thick curve is the past event horizon.
$E_1$ is the original bifurcation surface. $E_2$ is the location where the right and left future horizons cross. The magenta curve is a null ray that passes through wormhole, deviating to right boundary.
\label{fig:4.1}}
\end{figure}

However, after the time $t_\text{trav}$ on the boundary, the bifurcation surface ($E_1$ of Fig. \ref{fig:4.1}) is no longer spacelike to the AdS boundary, and one cannot define a bulk entanglement wedge using the surface $E_1$. This would render $S_\text{bulk}$ ill-defined. The resolution is to use the \emph{quantum extremal} surface. 

 More generally, it was proposed in \cite{EngelhardtWall} (and proven in \cite{DongLewkowycz}) that, at general orders in $1/N$, one should consider the entropy outside the quantum extremal surface, obtained by extremizing the total generalized entropy $S_\mathrm{gen} = A/4G_N + S_\text{bulk}$.  When calculating the ${\cal O}(1)$ piece of the entropy, these two prescriptions agree on the value of the entropy but \cite{EngelhardtWall,Dong:2016eik} argued that the location of the quantum extremal surface is also physically important, because it provides a natural boundary for how much of the bulk can be reconstructed from the CFT state on a single boundary.  One useful constraint on the location quantum extremal surface is the GSL, which states that $S_\mathrm{gen}$ is nondecreasing on any future horizon.

On a Cauchy slice prior to the time when the interaction is turned on, the geometry and bulk quantum state are that of the Hartle-Hawking state.  Thus the quantum (and classical) extremal surface is located at the bifurcation surface of the original black hole ($E_1$ of Fig. \ref{fig:4.1}).  On the other hand, after the interaction is over, the bulk quantum state of the fields changes and thus the quantum extremal surface must move.  By left-right symmetry of the spacetime (together with the fact that the joint state of the entire system is pure so that $S_{EE}$ is the same on both sides) it can it can only move along the vertical axis of symmetry of the spacetime.  Also, the GSL implies that the new location must be on or behind the causal horizon \cite{EngelhardtWall}, because otherwise it lies on a future horizon whose $S_\mathrm{gen}$ is generically increasing.

In fact, at first order in $h$, the GSL implies that the quantum extremal surface must lie exactly at the point $E_2$ in Fig. \ref{fig:4.1}, where the two future horizons intersect.  For since the GSL is true in every state \cite{Wall2012}, and saturated for the Hartle-Hawking state, it must also be saturated for any first order perturbation to the Hartle-Hawking state \cite{Wall:2009wm}.  But if $S_\mathrm{gen}$ is stationary along two linearly independent normal directions of $E_2$, then it must be a quantum extremal surface.

As noted above, the area and bulk entanglement entropies are identical for $E_1$ and $E_2$ at linear order in $h$ (as long as they are well-defined). Any effects arising from differences between $E_1$ and $E_2$ are suppressed by additional powers of $h$.

At second order in $h$, the GSL should no longer be saturated on the future horizon.  Hence $S_\text{bulk}$ is increasing with time at $E_2$, and the quantum extremal surface will instead be located slightly above the point $E_2$.

We have not followed the evolution of the quantum extremal surface at intermediate times, but it seems that it must gradually move upwards from $E_1$ to its final location above $E_2$.  After the interaction is over the boundary evolution is unitary, and hence neither $S_{EE}$ nor the quantum extremal surface changes.

\cite{EngelhardtWall} argued that the quantum extremal surface should always be spacelike to its corresponding CFT region.  In a sense this continues to be true, since $E_1$ is spacelike to all the boundary points prior to turning on the interaction, while $E_2$ is spacelike to all the points after the interaction is turned off.  But neither one is spacelike to the entire boundary for all time.  For example, a unitary operator applied to the right boundary at sufficiently early times might affect the value of $S_\mathrm{gen}(E_2)$, and hence the right CFT entropy after the interaction.  But that does not contradict any of the properties of the right CFT, since it does not have unitary time evolution (independent of the left CFT) during the period of the interaction.

Note that, if we assume that our holographic entropy prescription is correct when the CFT's are not coupled, it must necessarily also be correct when the CFT's are coupled.  Before the interaction is turned on, we can simply consider the Hartle-Hawking spacetime as if there were no interaction.  Similarly, after the interaction is over, we can consider a new spacetime which is dual to extrapolating the final state backwards in time, without any interaction.  Neither of these spacetimes corresponds to a traversable wormhole, but they can be used for purposes of calculating $S_{EE}$ before or after the interaction is turned on.  It is only when these two spacetimes are patched together, that they are seen to be a traversable wormhole geometry.

\section{Discussion}

We have demonstrated that the Einstein-Rosen bridge of a BTZ black hole becomes
slightly traversable after the addition of a two-boundary coupling.  (We expect that a similar effect also occurs in $D > 3$ bulk spacetime dimensions, although it is harder to calculate the exact form of the stress-tensor.)

From (\ref{eq:3.17}), we see that the integral $\int dUT_{UU}$, giving the
deviation of null rays from the horizon, is proportional to $h$, which implies that the wormhole opens up by an amount (in units where $\hbar = 1$)
\begin{equation}
\Delta V \sim \frac{h G_N}{R^{D-2}}
\end{equation}
where $\Delta V$ is the difference of $V$ coordinate between the future horizon and the first lightray which can get through the wormhole (see Fig. \ref{fig:4.1}), and we assume that the black hole radius $r_h$, the AdS length $L_{AdS}$, and the amount of time $\Delta t$ the interaction is turned on for are all of the same characteristic length scale $R$.

The wormhole is only open for a small proper time in the interior region.  This is quite different from the usual static wormhole solutions which do not have event horizons (e.g. \cite{Morris1988a}).  Nevertheless, radial lightrays originating on the boundary at arbitrarily early times will cross through the portal to the other side; in this sense the wormhole is open at arbitrarily early boundary times on either side.

A (test) astronaut
from one boundary can only go through the wormhole before it closes,
and she reaches the other boundary long after  the boundary-boundary interaction is turned on. One should note that since the coupling
we add breaks the Killing symmetry $H_{L}-H_{R}$, there is no
way to boost her back to a time before she entered the worm hole. Thus the way we glue the two boundaries fixes the relative time coordinate
between them, excluding the possibility of having closed time-like curves \cite{Morris1988}. Note that the traversable throat size depends
on the strength of the coupling and a signal transmitted through the
wormhole is only received at the other end after a very long time delay
if the gravitational effects of the coupling are small.  Furthermore, the thermofield double state that we require is an extremely fine tuned state, so it would be very difficult to prepare such a configuration in which the astronaut could enter at early times from the left.

We have not yet considered the backreaction on the geometry coming from an actual (non-test) astronaut traversing the wormhole throat.  An object travelling at light speed from left to right contributes to $T_{VV}$ but not to $T_{UU}$, so at the level of linearized gravity it will prevent objects from traversing in the \emph{other direction} (i.e. from right to left) but it will have no tendency to close the wormhole in the same direction that it is travelling.  This suggests that the objects can still traverse the wormhole even after taking into account their own gravitational back-reaction.\footnote{Presumably there is \emph{some} limit on how much information can get through, since the black hole on the other side cannot radiate more energy than its initial mass, but determining the precise limit would require going beyond the linearized regime.  There might also be an interesting limit on the total amount of \emph{information} which can get through the wormhole, coming from the Bousso bound \cite{Bousso:1999xy} (see \cite{Flanagan2000} for a proof in classical case) or its quantum generalization \cite{Bousso:2015mna,Strominger:2003br}.}

Another question concerns the interaction of the astronaut with the negative energy pulse of radiation travelling in the other direction.  In the frame of reference defined by Kruskal coordinates, a quantum traversing the wormhole must be blueshifted up to a frequency $1/{\Delta V}$, while the pulse coming in the other direction has a frequency of order $1/R$.  Here we are assuming that the interaction is turned on for about one light-crossing time $R$, and that there is no other time scale of relevance in the problem.  Although an incoming pulse with negative total energy is not allowed in classical scattering problems, we will nevertheless attempt to build intuition by comparing the situation to a normal field theory scattering problem.  The center-of-mass energy scale of the collision is given by
\begin{equation}
\sqrt{s} \sim \sqrt{\frac{R^{D-4}}{h G_N} }.
\end{equation}
Since $G_N \sim L_\text{planck}^{D-2}$, the center-of-mass energy is below the Planck scale in $D = 3$ (i.e. a BTZ black hole with any extra dimensions compactified at the Planck scale) but not when $D > 3$.  However, even in higher dimensions we do not expect that full quantum gravity effects will be important.   We nevertheless expect that it is legitimate to use the eikonal approximation, in which one solves for the propagation of each particle on the background field generated by the other particle.  This corresponds to resumming ladder Feynman diagrams, whose amplitude scales with various powers of
\begin{equation}
\frac{G_N s}{b^{D-4}}\sim h^{-1},
\end{equation}
where $b$ is the impact parameter, and we have used the fact that $b \sim R$ (except for small tails of the wavefunction).  Non-eikonal Feynman diagrams should be suppressed by additional powers of $G_N$ relative to eikonal diagrams with the same $s$ dependence \cite{Amati:1993tb}.  Therefore we can consistently consider scenarios in which only the eikonal scattering is relevant, in which our calculation of the geometry shows that the wormhole is traversable if the particle is sent in before a time $\Delta t \approx R \ln (R / h L_\text{planck})$ prior to the interaction being turned on. However if the particle is sent in more than $\frac{3}{2}\Delta t$ prior to the interaction time, then the eikonal approximation breaks down and there may be large back-reaction, invalidating our analysis. When $h \sim 1$, these times are of order the scrambling time.

It is interesting to consider what would happen if the two black holes were in the same component of space, rather than in different asymptotic regions.  If the black holes were in a suitably entangled state, they should be connected by an Einstein-Rosen bridge \cite{Maldacena2013}, with the QFT state near the horizon close to the Hartle-Hawking state.  The direct boundary interaction could then be produced by propagation through the ambient spacetime---this would be the same as the interaction we studied, except with a time delay.  A similar calculation would then lead to a traversable wormhole.  The negative ANE could be understood as coming from the Casimir effect associated to the cycle in space going from one black hole to the other in the ambient space and then threading the wormhole.  Of course, the effect would be enhanced if the signals sent between the black holes were directed and amplified (otherwise the Casimir energy would be extremely tiny if the black holes were far apart).  No causal paradoxes would arise because the traversability depends on backreaction due to the existence of a casual path between the black holes in the ambient spacetime.

Since any infinite null geodesic which makes it through a wormhole must be chronal (as discussed in the Introduction), such wormholes do not enable one to travel faster than light over long distances through space.  Hence traversable wormholes are like getting a bank loan: you can only get one if you are rich enough not to need it.

The traversable wormhole we found has an interesting interpretation in the context of ER=EPR \cite{Maldacena2013}.
Maldacena and Susskind conjectured that any pair of entangled 
quantum systems are connected by an Einstein-Rosen bridge (the non-traversable
wormhole).
The crucial difference in our work is that
we allow interaction between the entangled systems, which is 
assumed to be negligible in ER=EPR.
What we have shown is that 
in this case the Einstein-Rosen bridge can open to become a traversable wormhole.

Our example thus provides a way to operationally verify a salient feature of ER=EPR that observers from opposite sides of an entangled pair of systems may meet in the connected interior. Since in \cite{Maldacena2013} any such meeting is trapped behind the horizon, it is not obvious how its occurrence could be confirmed by exterior or CFT measurements. What we found is that if, after the observers jump into their respective black holes, a boundary-boundary coupling is activated, then the Einstein-Rosen can be rendered traversable, and the meeting inside may be seen from the boundary. This seems to suggest that the ER=EPR wormhole connection was physically ``real''.  But since all measurements in the CFT description are governed by the rules of linear quantum mechanics, it seems like any explicit operational verification of the existence of the wormhole would also correspond to a linear quantum measurement.  It might be interesting to check the compatibility of these ideas with the linearity of measurements made behind the horizon, discussed in \cite{marolf2012eternal}.

What is the quantum information theory interpretation of such a traversable wormhole?  A curious feature of the transmission of a qubit, $Q$, through the wormhole is that it appears to be sent ``via the entanglement'', rather than directly by the inter-boundary coupling.  (Note that the traversable portion of the wormhole is close to the bifurcation point, which describes the subspaces of the left and right Hilbert spaces that are the most entangled in the thermofield double state.)  There are several ways to see that the quantum information of $Q$ is not simply being sent directly through the boundaries. First, the commutator of $Q$ (for example when it is first injected into the interior from the left boundary) with the interaction Hamiltonian is extremely small near the thermofield double state. Furthermore, at the time the interaction is activated, $Q$ is in fact spacelike separated from the boundary in the bulk picture, so in the bulk approximation $Q$ and ${\cal O}$ are independent quantum variables. From the CFT perspective, this is because $Q$ has thermalized into the left system before the ${\cal O}_L {\cal O}_R$ interaction is turned on, so no quantum information about $Q$ appears to be accessible to the operator ${\cal O}$. Of course, the boundary coupling is nevertheless crucial for the existence of the traversable wormhole.

This situation is somewhat analogous to what occurs in quantum teleportation.  Entanglement alone cannot be used to transmit information, and no qubit, $Q$, from the left can traverse the bridge to the right if the left and right systems are dynamically decoupled. However, if additional classical information is sent from the left to the right, a qubit can be transmitted - this is referred to as quantum teleportation.
Suppose Alice and Bob share a maximally entangled pair of qubits, $A$ and $B$. Alice can then transmit the qubit $Q$ to Bob by sending only the classical output of a measurement on the $Q$-$A$ system. Depending on which of the 4 possible results are obtained, Bob will perform a given unitary operation on the qubit $B$, which is guaranteed to turn it into the state $Q$.

Unlike the usual description of quantum teleportation, in our example it is essential that the channel between the left CFT, $A$, and the right CFT, $B$, is a quantum one. For example, if one projected onto eigenstates of the operator ${\cal O}_L$, then the configuration would simply look like a particular quantum state (the projection of $|\textrm{tfd}\rangle$) evolving under the decoupled Hamiltonians together with an action by a purely right unitary, which can never lead to a traversable wormhole.  This makes sense, because in the standard description of quantum teleportation, the measurement performed by Alice is a projection onto an eigenstate, which instantly results in the pattern of $Q$ being contained in the system $B$. This would not be described by a physical motion through the wormhole in the bulk. Teleportation in this sense has been discussed in the dual gravity language by \cite{susskind2016er, numasawa2016epr, marolf2012eternal}.

However, in the exact, fully quantum description of the quantum teleportation protocol, there is a particular dynamical process given by the unitary evolution $V = \sum_i P^{Q A}_i U^B_i$ that governs the transmission of the ``classical'' information and the subsequent appropriate transformation of a qubit in the $B$ quantum system. Here $P^{Q A}_i$ are a complete mutually exclusive set of projectors on the $Q$-$A$ system that describe Alice's measurement, and $U^B_i$ is the unitary transformation performed by Bob given the data $i$. The classical information transmitted from Alice to Bob was encoded by the index $i$.


Treating $V$ as a time dependent interaction Hamiltonian can result in negative ANE along the horizon if the original entanglement between $A$ and $B$ was well described by a large Einstein-Rosen bridge, which will render the wormhole traversable. This is a description in which the time scales and processes of decoherence and measurement by Alice are resolved, and treated as physical dynamical evolution. In such a ``microscopic'' description of quantum teleportation, the qubit $Q$ must physically evolve from the left to the right. Of course in the limit that Alice's measurement is essentially instantaneous and classical, the traversable window will be very small (and not well described by a semiclassical spacetime) - just enough to let the single qubit $Q$ pass through.
Therefore, we propose that the gravitational dual description of quantum teleportation understood as a dynamical process is that the qubit passes through the ER=EPR wormhole of the  entangled pair, $A$ and $B$, which has been rendered traversable by the required interaction.


Another possible interpretation of our result is to relate it to the recovery of information described in \cite{PreskillHayden}.  Assuming that black hole evaporation is unitary, it is in principle possible to eventually recover a qubit which falls into a black hole, from a quantum computation acting on the Hawking radiation.  Assuming that you have access to an auxiliary system maximally entangled with the black hole, and that the black hole is an efficient scrambler of information, it turns out that you only need a small (order unity) additional quantity of Hawking radiation to reconstruct the qubit.  In our system, the qubit may be identified with the system that falls into the black hole from the left and gets scrambled, the auxiliary entangled system is the CFT on the right, and the boundary interaction somehow triggers the appropriate quantum computation to make the qubit reappear again, after a time of order the scrambling time $R \ln (R / L_\text{planck})$.\footnote{We thank Juan Maldacena for suggesting this interpretation.}

\section*{Acknowledgements}

We thank Ofer Aharony, Daniel Harlow, Juan Maldacena, Sudipta Sarkar, Douglas Stanford and Andy Strominger for helpful and stimulating discussions.  DLJ and PG were supported in part by NSFCAREER grant PHY-1352084 and by a Sloan Fellowship. AW was supported by the Institute for Advanced Study, by the Martin A. and Helen Chooljian Membership Fund, and NSF grant PHY-1314311. AW is grateful for typing support from Geoff Penington.

\appendix

\section{$\int dUT_{UU}$\label{sec:intTUU}}

Using (\ref{eq:3.8}), the integrated null energy is
\begin{equation}
\int_{U_{0}}^{\infty}dUT_{UU}=-4h\D C_{0}\int_{U_{0}}^{\infty}dU\lim_{U'\ra U}\del_{U}G(U,U';U_{0}),\label{eq:a1}
\end{equation}
where
\begin{equation}
G(U,U';U_{0})\equiv\int_{U_{0}}^{U}dU_{1}\int_{1}^{\f U{U_{1}}}\f{dy}{\sqrt{y^{2}-1}}\f{U_{1}^{\D}}{(U-U_{1}y)^{\D}(U'U_{1}+y)^{\D+1}}.
\end{equation}
Note that
\begin{equation}
\lim_{U'\ra U}\del_{U}G(U,U';U_{0})=\del_{U}G(U,U;U_{0})-\del_{U}^{(2)}G(U,U;U_{0}),
\end{equation}
where $\del_{U}^{(2)}$ indicates a derivative with respect to the second variable.
(\ref{eq:a1}) becomes
\begin{equation}
\int_{U_{0}}^{\infty}dUT_{UU}=-4h\D C_{0}\left(G(\infty,\infty;U_{0})-G(U_{0},U_{0};U_{0})-\int_{U_{0}}^{\infty}dU\del_{U}^{(2)}G(U,U;U_{0})\right)\label{eq:a4}
\end{equation}
By (\ref{eq:3.9}), and changing to the $z$ variable,
\begin{align}
G(U,U;U_{0}) & \propto\int_{U_{0}}^{U}dU_{1}\f{U_{1}^{\D-1/2}}{(U-U_{1})^{\D-1/2}(1+UU_{1})^{\D+1}}F_{1}(\f 12;\f 12,\D+1;\f 32-\D;\f{U_{1}-U}{2U_{1}},\f{U_{1}-U}{U_{1}(1+UU_{1})})\nonumber \\
 & =\int_{0}^{1}dz\f{((U-U_{0})z+U_{0})^{\D-1/2}F_{1}(\f 12;\f 12,\D+1;\f 32-\D;-\f{(U-U_{0})(1-z)}{2((U-U_{0})z+U_{0})},-\f{(U-U_{0})(1-z)}{((U-U_{0})z+U_{0})(1+U((U-U_{0})z+U_{0}))})}{(U-U_{0})^{\D-3/2}(1-z)^{\D-1/2}(1+U((U-U_{0})z+U_{0}))^{\D+1}}
\end{align}
which immediately implies $G(U_{0},U_{0};U_{0})=0$ given that $\D<3/2$.
For the large $U$ limit, $G(\infty,\infty;U_{0})$,  the prefactor
of $F_{1}$ above decays at least as fast as $U^{-\D}$ and the $F_{1}$
part becomes
\begin{equation}
F_{1}(\f 12;\f 12,\D+1;\f 32-\D;-\f{1-z}{2z},0)={}_{2}F_{1}(\f 12,\f 12;\f 32-\D;\f{z-1}{2z})=\left(\f{2z}{z+1}\right)^{1/2}{}_{2}F_{1}(\f 12,1-\Delta;\f 32-\D;\f{1-z}{1+z})
\end{equation}
which leads to
\begin{equation}
G(\infty,\infty;U_{0})\sim U^{-\D}\int_{0}^{1}dz\left(\f{2z}{z+1}\right)^{1/2}(1-z)^{-\D+1/2}{}_{2}F_{1}(\f 12,1-\Delta;\f 32-\D;\f{1-z}{1+z})\ra0
\end{equation}
where in the last step we used the fact that the $z$ integral is
finite due to the property of hypergeometric function:
\begin{equation}
_{2}F_{1}(\f 12,1-\Delta;\f 32-\D;0)=1,\qquad\lim_{z\ra0}z^{1/2}{}_{2}F_{1}(\f 12,1-\Delta;\f 32-\D;\f{1-z}{1+z})\sim z^{1/2}\log\f{2z}{z+1}\ra0
\end{equation}

The integral of $T_{UU}$ is simplified to be
\begin{align}
 & -\f 1{4h\D C_{0}}\int_{U_{0}}^{\infty}dUT_{UU}\nonumber \\
= & -\int_{U_{0}}^{\infty}dU\int_{U_{0}}^{U}dU_{1}\int_{1}^{\f U{U_{1}}}\lim_{U'\ra U}\del_{U'}\f{dy}{\sqrt{y^{2}-1}}\f{U_{1}^{\D}}{(U-U_{1}y)^{\D}(U'U_{1}+y)^{\D+1}}\nonumber \\
= & \int_{U_{0}}^{\infty}dU_{1}\int_{U_{1}}^{\infty}dU\int_{1}^{\f U{U_{1}}}\f{dy}{\sqrt{y^{2}-1}}\f{(\D+1)U_{1}^{\D+1}}{(U-U_{1}y)^{\D}(UU_{1}+y)^{\D+2}}\nonumber \\
= & \int_{U_{0}}^{\infty}dU_{1}\int_{U_{1}}^{\infty}dU\f{(\D+1)\G(\f 12)\G(1-\D)U_{1}^{2\D+3}(U+U_{1})^{-1/2}}{\Gamma(\f 32-\D)(U-U_{1})^{\D-1/2}U^{\D+2}(1+U_{1}^{2})^{\D+2}}F_{1}(1-\D;\f 12,\D+2;\f 32-\D;\f{U-U_{1}}{U+U_{1}},\f{U-U_{1}}{U(1+U_{1}^{2})})\label{eq:a9}
\end{align}
For further simplification, we use (\ref{eq:3.17-1}) and define $w=\f{U-U_{1}}{U+U_{1}}$
to rewrite (\ref{eq:a9}) as
\begin{align}
 & -\f 1{4h\D C_{0}}\int_{U_{0}}^{\infty}dUT_{UU}\nonumber \\
= & \sum_{m}\f{(\D+1)\G(\f 12)\G(1-\D)(1-\D)_{m}(\D+2)_{m}2^{m+1-\D}}{m!\Gamma(\f 32-\D)(\f 32-\D)_{m}}\int_{U_{0}}^{\infty}dU_{1}\f{U_{1}^{2}}{(1+U_{1}^{2})^{\D+m+2}}\nonumber \\
 & \times\int_{0}^{1}dw\f{w^{m-\D+1/2}(1-w)^{2\D}}{(1+w)^{\D+m+2}}{}_{2}F_{1}(1-\D+m;\f 12;\f 32-\D+m;w)\nonumber \\
= & \sum_{m}\f{(\D+1)\G(\f 12)\G(1-\D)(1-\D)_{m}(\D+2)_{m}2^{m+1-\D}}{m!\Gamma(\f 32-\D)(\f 32-\D)_{m}}\int_{U_{0}}^{\infty}dU_{1}\f{U_{1}^{2}}{(1+U_{1}^{2})^{\D+m+2}}\nonumber \\
 & \times\f{\G(\f 32-\D+m)\G(2\D+1)^{2}}{2^{\D+m+2}\G(\f 32+2\D)\G(2+\D+m)}{}_{2}F_{1}(2\D+1,2\D+1;\f 32+2\D;\f 12)\nonumber \\
= & \sum_{m}\f{\G(\f 12)\G(1-\D)\G(2\D+1)^{2}(1-\D)_{m}}{m!\Gamma(\f 32+2\D)\G(\D+1)2^{2\D+1}}{}_{2}F_{1}(2\D+1,2\D+1;\f 32+2\D;\f 12)\nonumber \\
 & \times\f{(U_{0}^{-2})^{\D+m+1/2}}{2(\D+m+1/2)}{}_{2}F_{1}(\f 12+\D+m,2+\D+m;\f 32+\D+m;-U_{0}^{-2})\nonumber \\
= & \f{\G(\f 12)\G(1-\D)\G(2\D+1)^{2}}{2^{2\D+2}(\D+\f 12)\Gamma(\f 32+2\D)\G(\D+1)}\f 1{(1+U_{0}^{2})^{\D+1/2}}{}_{2}F_{1}(2\D+1,2\D+1;\f 32+2\D;\f 12)\nonumber \\
 & \times\sum_{m}\f{(1-\D)_{m}(\f 12+\D)_{m}}{m!(\f 32+\D)_{m}}\left(\f 1{1+U_{0}^{2}}\right)^{m}{}_{2}F_{1}(\f 12+\D+m,-\f 12;\f 32+\D+m;\f 1{1+U_{0}^{2}})\nonumber \\
= & \f{\pi\G(1-\D)\G(2\D+1)^{2}}{2^{2\D+2}(\D+\f 12)\G(\D+1)^{3}}\f{_{2}F_{1}(\f 12+\D,\f 12-\D;\f 32+\D;\f 1{1+U_{0}^{2}})}{(1+U_{0}^{2})^{\D+1/2}}
\end{align}
where in fifth line we used \cite{Prudnikov1992}
\begin{align}
 & \int_{0}^{y}\f{x^{c-1}(y-x)^{\b-1}}{(1-zx)^{\r}}{}_{2}F_{1}(a,b;c;\f xy)dx\nonumber \\
= & \f{y^{c+\b-1}}{(1-yz)^{\r}}\f{\G(c)\G(\b)\G(c-a-b+\b)}{\G(c-a+\b)\G(c-b+\b)}{}_{3}F_{2}(\b,\r,c-a-b+\b;c-a+\b,c-b+\b;\f{yz}{yz-1})\nonumber \\
 & [y,\Re c,\Re\b,\Re(c-a-b+\b)>0;|\arg(1-yz)|<\pi]
\end{align}
in sixth line we used
\begin{equation}
\int_{b}^{\infty}dx\f{x^{2}}{(1+x^{2})^{a}}=\f{b^{-2a+3}}{2a-3}{}_{2}F_{1}(a-\f 32,a;a-\f 12;-b^{-2})
\end{equation}
and in last step we used \cite{Prudnikov1992}
\begin{align}
\sum_{k=0}^{\infty}\f{(a)_{k}(b')_{k}}{k!(c)_{k}}x^{k}{}_{2}F_{1}(a+k,b;c+k;x) & ={}_{2}F_{1}(a,b+b';c;x)\\
_{2}F_{1}(2\D+1,2\D+1;\f 32+2\D;\f 12) & =\f{\pi^{1/2}\G(\f 32+2\D)}{\G(1+\D)^{2}}
\end{align}

In the end, restoring $\ell$, we find the following relatively simple result
\begin{equation}
\int_{U_{0}}^{\infty}dUT_{UU}=-\f{h\G(2\D+1)^{2}}{2^{4\D}(2\D+1)\Gamma(\D)^{2}\G(\D+1)^{2}\ell}\f{_{2}F_{1}(\f 12+\D,\f 12-\D;\f 32+\D;\f 1{1+U_{0}^{2}})}{(1+U_{0}^{2})^{\D+1/2}}
\end{equation}
If we turn off the interaction at $U_{f}$, the integral is just the
difference between $\int_{U_{0}}^{\infty}dUT_{UU}$ and $\int_{U_{f}}^{\infty}dUT_{UU}$.

\bibliographystyle{plain}
\bibliography{citedate}

\end{document}